\documentclass[pdftex,twocolumn,epjc3]{svjour3}          

\RequirePackage[T1]{fontenc}
\smartqed  

\RequirePackage{graphicx}
\RequirePackage{mathptmx}      
\RequirePackage{flushend}
\RequirePackage[numbers,sort&compress]{natbib}
\RequirePackage[colorlinks,citecolor=blue,urlcolor=blue,linkcolor=blue]{hyperref}
\usepackage{amsmath,amssymb,physics}

\DeclareMathOperator{\AdS}{AdS}
\DeclareMathOperator{\SO}{SO}
\newcommand*{\half}{\frac{1}{2}}
\newcommand*{\sixth}{\frac{1}{6}}

\begin{document}

\title{Perturbations to Kink-Like Topological Defects in
2D Anti de Sitter Spacetime
}


\author{Orlando Alvarez\thanksref{e1,addr1}
        \and
        Matthew Haddad\thanksref{e2,addr1} 
}

\thankstext{e1}{e-mail: oalvarez@miami.edu}
\thankstext{e2}{e-mail: m.haddad@miami.edu}

\institute{Department of Physics, University of Miami, 1320
Campo Sano Ave, Coral Gables, FL 33146, USA\label{addr1}
}

\date{Received: date / Accepted: date}

\maketitle

\begin{abstract}
We examine codimension--1 topological defects whose associated
worldline is geodesically embedded in $\AdS_{2}$.  This discussion
extends a previous study of exact analytical solutions to the
equations of motion of topological defects in $\AdS_{n}$ in a
particular limit where the masses of the scalar and gauge field
vanish.  We study the linear perturbations about the zeroth order
kink-like solution and verify that they are stable.  We also discuss
general features of the perturbation expansion to all orders.
\end{abstract}

	\section{Introduction}
This is the second in an ongoing series of articles in which we study
the solutions to the equations of motion of topological defects in
anti de Sitter ($\AdS$) spacetime.  In our previous
paper~\cite{alvarezHaddad2018}, we were able to show that there exist
exact analytical solutions to the equations of motion for a
topological defect in $\AdS_{n}$ in what we dubbed the ``double BPS
limit'' owing to the similarity of our approach with the parameter
limit used by Bogomolny, Prasad, and Sommerfield (BPS)
\cite{bogomolny1976,prasadSommerfield1975}.  There are other studies
on topological defects in $\AdS$, such as the work of Lugo et
al.~\cite{lugo1999,lugo2000} on magnetic monopoles in $\AdS_{4}$ and
more recently the work of Ivanova et al.~\cite{ivanova2017} on
finite-energy Yang-Mills field theory in $\AdS_{4}$.  Kink-like
defects in flat space are discussed by Coleman in his famous Erice
lectures \cite{coleman1985}.  Manton and Sutcliffe give a good
treatment of topological defects in general in their book
\cite{mantonSutcliffe2004}.

There is a long history of potential physical consequences of
topological defects such as magnetic monopoles in spacetime, for a
review see \cite{magneticMonopoles}.  Also, there are many studies of
the physics of cosmic strings~\cite{Vachaspati:2015} where the
spacetime background is taken to be flat for simplicity, or the length
scale of the string is much smaller than the radii of curvature of the
spacetime.  We decided to search for exact topological solitons in
anti de Sitter space which like Minkowski space is a maximally
symmetric Lorentzian manifold.  We were motivated by the observation
that $\AdS$ has a tautological length scale given by its finite radius
of curvature $\rho$, and you can study solutions to the equation of
motion where this is the dominant length scale.  This is not possible
in Minkowski space.  The solutions we discovered in
\cite{alvarezHaddad2018} are of this type, and in this article we
study perturbations in which the other length scale, determined by the
flat space mass of the scalar field, starts to become applicable.  We
note that in $\AdS_{2}$, for two spacelike separated point a proper
distance $\nu$, the two point correlation function, see
\cite[eq.~(4.6)]{alvarezHaddad2018}, decays exponentially as
$e^{-m_{*}\nu}$ where
\begin{equation*}
m_{*} = \frac{1}{2\rho} + \sqrt{m_{\phi}^{2} +
\left(\frac{1}{2\rho}\right)^{2}},
\end{equation*}
and $m_{\phi}$ is the flat space mass of the scalar field.  Even if
$m_{\phi}=0$, the limit we considered in our previous paper, we have
an effective finite correlation length of $\rho$ for the decay of the
two point scalar field function.  Our solution is a kink that lives in
the spatially infinite manifold $\AdS_{2}$ but its energy density is
localized in a spacelike region roughly of size $\rho$.  The spirit of
our solutions is different than the traditional discussions of
topological defects in spacetime where correlation length of the field
theory may be large but it is still very much smaller than the radii
of curvature of the spacetime.  We consider topological defects that
are of cosmological size, but at this moment we do not know any
physical applications.  You can wildly speculate that the localization
of the energy of our topological defects might lead to cosmological
structure formation, such as dark matter aggregation, over a distance
scale $\rho$.  We not studied this problem but we mention that we also
found  Nielsen-Olesen string solutions that may more useful for
cosmological applications than domain walls (the kink).

We briefly explain some notational conventions.  Our work concerns
analytical solutions for $\SO(l)$ Higgs field theories in a static
$\AdS_{n}$ background.  These admit maximally symmetric topological
defects of dimension $p$.  The world brane associated with one of
these defects is then a $q=p+1$ dimensional timelike totally geodesic
submanifold $\Sigma^{q} \approx \AdS_{q}$ embedded in $\AdS_{n}$.  The
value of $l$ is the number of transverse dimensions, $n = q + l$.  Our
framework allows us to seek exact analytical solutions for defects
categorized by pairs $(q,l)$, where $q\geq 1$ and $l \geq 1$.  In this
article we study the one-dimensional kink that is the $(1,1)$ defect.
In a forthcoming article, we generalize this to the case of $(q,1)$
dimensional defects in $\AdS_{q+1}$.  Due to the curvature of the
background $\AdS_{q+1}$, the equations of motion for the topological
kink depend on the value of $q$; a very different situation than in
Minkowski space.
	
We have mentioned the double BPS limit thus far without much
explanation.   BPS studied monopole solutions to
the equations of motion for topological defects in $\mathbb{R}^{3+1}$ in
the limit that the mass of the scalar field vanishes while preserving
some boundary conditions.  They found that there were exact analytical
solutions.  This work was done in flat space, where the length scale
was set by the gauge field mass.  In non-flat constant curvature
spaces, the radius of curvature $\rho$ sets a third length scale.
Therefore, one can take the limit in the equations of motion where the
masses of both the scalar and the gauge fields vanish and obtain
equations that can have exact analytical solutions with a single
length scale $\rho$.  This is the double BPS limit.  For a more
thorough explanation of this process, we direct the reader to our
previous article~\cite{alvarezHaddad2018}.
	
The double BPS equations of motion do not follow directly from an
action, but rather from a limit imposed on the equations of motion for
a topological defect in curved spacetime.  To this end, the double BPS
solution is a starting point to a perturbative solution to these
original equations in which the mass of the fields are taken to be on
the same order as the perturbation.  We pursue such an approximate
solution in this article, and show that the linear
perturbations are stable, and uniquely determine the second, third, and higher
order corrections.  A motivation for studying these corrections
stems from a pursuit of greater understanding of the original
solution.  Since we have obtained a topologically stable solution to
the equations of motion, any small perturbation in the mass should
produce real eigen-frequencies.  In this article, we seek to
verify this for the specific case of the kink defect in $\AdS_{2}$ and
to prime the reader for the discussion on the stability of the more
general $\AdS_{n}$ kink defect in a forthcoming article.  

There is an alternative approach we did not pursue in this paper.  We
could do perturbation about the static numerical solution to the exact
kink equation~\eqref{eq:eomNoDimConformal}.  We did not follow this
approach because we wanted to exploit the existence of an exact kink
solution in the double BPS limit, and use this simplest case of
$\AdS_{2}$ as a testbed for studying the kink in $\AdS_{n}$.

\section{Deriving the Equation of Motion}
\subsection{The Conformal Metric}
In our previous paper, we discussed the procedure for writing the
action and deriving the equations of motion for a
spherically-symmetric topological defect in $\AdS_{n}$
\cite{alvarezHaddad2018}.  In that discussion, we adopted an
$\SO(l)$-gauged Higgs model embedded into $\AdS_{n}$, where $n \geq
2$.  We restrict our discussion to maximally symmetric defects, so
that $n = q + l$ where $q$ is the dimension of the topological defect
we are studying.  In this article we focus on the $(1,1)$ defect,
which is the kink-like defect that extends from $\AdS_{1}$ out into
$\AdS_{2}$.  In this case, there is only a scalar field $\phi$.  The
worldline associated with the defect is $\Sigma^{1} \approx \AdS_{1}$.
This is a one-dimensional timelike submanifold, a timelike curve, and
so we simply take the metric as $\dd{s}_{\AdS_{1}}^{2} =
-\dd{\tau}^{2}$.  If $\rho$ is the radius of curvature of $\AdS_{2}$
and $\nu$ as the signed distance to a point to $\Sigma^{1}$, we can
write the metric of $\AdS_{2}$ as
\begin{align}
\dd{s}_{\AdS_{2}}^{2} &= 
\cosh[2](\frac{\nu}{\rho})\dd{s}_{\AdS_{1}}^{2} + \dd{\nu}^{2} 
\nonumber \\
&= -\cosh[2](\frac{\nu}{\rho}) \dd{\tau}^{2} + \dd{\nu}^{2} 
\label{eq:initialMetric}
\end{align}
where $-\infty < \tau < +\infty$ and $-\infty < \nu < +\infty$.

The Lagrangian for the model is
\begin{equation}
\mathcal{L} = -\frac{1}{2}\partial_{\mu}\phi \partial^{\mu}\phi - 
\frac{\lambda}{8}\qty(\phi^{2} - \phi_{0}^{2})^{2} 
\label{eq:lagrangian}
\end{equation}
where $\lambda$ is the scalar field coupling and $\phi_{0}$ is the 
field's vacuum expectation value at spatial infinity (that is, $\phi 
\xrightarrow{\nu \to \pm \infty} \phi_{0}$).

The action is then
\begin{align}
I &= \int_{-\infty}^{\infty} \int_{-\infty}^{\infty} 
\sqrt{-g}\,\mathcal{L} \nonumber \\
&= -\int_{-\infty}^{\infty} \int_{-\infty}^{\infty} \dd{\tau}
\dd{\nu} 
\cosh(\frac{\nu}{\rho}) \qty[\frac{1}{2} \partial_{\mu}\phi 
\partial^{\mu}\phi + \frac{\lambda}{8}\qty(\phi^{2} - 
\phi_{0}^{2})^{2}] \nonumber \\
 &= -\int_{-\infty}^{\infty} \dd\tau
\int_{-\infty}^{\infty} 
\dd{\nu} 
\cosh(\frac{\nu}{\rho}) 
\biggl[-\frac{1}{2\cosh[2](\nu/\rho)}\qty(\pdv{\phi}{\tau})^{2} 
\nonumber \\
&\quad + 
\frac{1}{2}\qty(\pdv{\phi}{\nu})^{2} +
\frac{\lambda}{8}\qty(\phi^{2} 
- \phi_{0}^{2})^{2}\biggr] \label{eq:action}
\end{align}
For ease of computation we will scale everything by the radius of
curvature in this article, so that we can put everything in terms of
dimensionless variables.  This amounts to the substitutions $\nu \to
\rho \nu$, $\tau \to \rho \tau$ and $\phi \to \phi_{0} \phi$.  We also
take note of the combination $\lambda \phi_{0}^{2}$ which is the
flat-space mass squared, $m_{\phi}^{2}$, of the field and then define
the dimensionless mass $\mu^{2} = (m_{\phi}\rho)^{2}$.  The action
\eqref{eq:action} gives the equation of motion that was studied in
\cite{alvarezHaddad2018}
\begin{multline}
-\pdv[2]{\phi}{\tau} 
+ \cosh[2](\nu) \biggl[\pdv[2]{\phi}{\nu} 
\\+
\tanh(\nu) \pdv{\phi}{\nu} - \frac{1}{2}\mu^{2}\phi (\phi^{2}-1) 
\biggr]
 = 0
\label{eq:eomNoDim}
\end{multline}

In this $(1+1)$ dimensional case, the linear perturbation analysis is
better done in coordinates where the metric is conformal to the flat
space metric.  We make the substitution $x = \arctan(\sinh(\nu))$
which gives us the metric
\begin{align}
\dd{s}^{2} &= \frac{\rho^{2}}{\cos^{2} x}\; \qty(-\dd{\tau}^{2} +
\dd{x}^{2})\,.
\label{eq:metricConformal}
\end{align}
This shows that $\AdS_{2}$ is conformally equivalent to the strip
$\mathbb{R} \times \qty[-\frac{\pi}{2}, \frac{\pi}{2}] \subset
\mathbb{M}^{2}$. The action in these coordinates is then
\begin{align}
I &= -\phi_{0}^{2} \int_{-\infty}^{\infty} \int_{-\pi/2}^{\pi/2}
\dd{\tau} \dd{x} \biggl[\frac{1}{2} \biggl\{-\qty(\pdv{\phi}{\tau})^{2} +
\qty(\pdv{\phi}{x})^{2}\biggr\} 
\nonumber \\
&\quad + \frac{1}{8}\mu^{2} \sec[2](x)
\qty(\phi^{2} -
1)^{2}\biggr], \label{eq:actionConformal}
\end{align}
with equation of motion
\begin{align}
-\pdv[2]{\phi}{\tau} + \pdv[2]{\phi}{x} &=
\frac{1}{2} \mu^{2} \sec[2](x) \phi \qty(\phi^{2} - 1).
\label{eq:eomNoDimConformal}
\end{align}
We analyze the linearization of this equation in this paper.  Note
that the differential operator on the left-hand side is just the
two-dimensional flat space d'Alembertian.  This equation admits a
static solution but we have not found an explicit analytical
expression.  This static nonlinear ODE is easily solved numerically
and a solution is plotted in figure~\ref{fig:phiNumerical}.

\subsection{The Double BPS Solution}
In our previous article, we noted the existence of an exact analytical
solution to \eqref{eq:eomNoDim} in the limit where $\mu \to 0$.  This
solution is possible because of the nonzero radius of curvature of
$\AdS_{2}$ and we refer to it as the ``double BPS'' solution due to
the similarity of this approach to that of Bogomolny, Prasad, and
Sommerfield \cite{bogomolny1976,prasadSommerfield1975} in the
flat-space case.  In the double BPS limit using the conformal
coordinates, the equation of motion for a static defect is
\begin{align}
\dv[2]{\varphi_{0}}{x} = 0, \label{eq:eomDbpsConformal}
\end{align}
with boundary conditions $\varphi_{0}\qty(\pm \frac{\pi}{2}) =
\pm 1$.
The solution to this ODE is simply
\begin{align}
\varphi_{0}(x) = \frac{2}{\pi}\, x \label{eq:dbpsSolnConformal}.
\end{align}
	
\subsection{Introducing the Small-Mass Perturbation}
We now use \eqref{eq:dbpsSolnConformal} as the starting point for a
perturbative solution where the small perturbation parameter is the
mass term.  The perturbation expansion to all orders is discussed in
Section~\ref{sec:GreensFn}.  Here we only consider the first two terms
of the expansion and write the field as $\phi(\tau,x) =\varphi_{0}(x)
+ \epsilon \varphi_{1}(\tau,x) + \epsilon^{2} \varphi_{2}(\tau,x) +
\order{\epsilon^{3}}$ with $\abs{\epsilon} \ll 1$.  We now insert this
into \eqref{eq:eomNoDimConformal} and keep only the terms to
$\order{\epsilon^{2}}$:
\begin{multline}
\dv[2]{\varphi_{0}}{x} +
\epsilon\qty( -\pdv[2]{\varphi_{1}}{\tau} + \pdv[2]{\varphi_{1}}{x})
+ \epsilon^{2} \qty(-\pdv[2]{\varphi_{2}}{\tau} +
\pdv[2]{\varphi_{2}}{x})   \\
= \frac{1}{2} \mu^{2} \sec[2](x)
\Big[\varphi_{0}
\qty(\varphi_{0}^{2} - 1) + \epsilon \varphi_{1} \qty(3 \varphi_{0}^{2} - 1) \\
 + \epsilon^{2}
\qty(3\varphi_{0} \varphi_{1}^{2} + \varphi_{2} (3\varphi_{0}^{2} -
1)) \Big]
\label{eq:eomPert}
\end{multline}

Notice that $\varphi_{0}(x)$ is a linear function, so its second
derivative vanishes.  Now we choose the mass term to be of the same
order as the perturbation, $\mu^{2} = \epsilon \tilde{\mu}^{2}$.  We are left
with the differential equations for the first- and second-order
perturbations:
\begin{align}
-\pdv[2]{\varphi_{1}}{\tau} + \pdv[2]{\varphi_{1}}{x} &= \frac{1}{2}
\tilde{\mu}^{2} \sec[2](x) \varphi_{0} \qty(\varphi_{0}^{2} - 1)
\label{eq:eomPertFirstOrderHom} \\
-\pdv[2]{\varphi_{2}}{\tau} + \pdv[2]{\varphi_{2}}{x} &= \frac{1}{2}
\tilde{\mu}^{2} \sec[2](x) \varphi_{1} \qty(3 \varphi_{0}^{2} - 1)
\label{eq:eomPertFirstOrder}
\end{align}
Since $\varphi_{0}$ satisfies the required boundary conditions, we
have that  perturbative expansion functions satisfy
$\varphi_{1}(\tau,\pm \pi/2) =0$ and $\varphi_{2}(\tau,\pm \pi/2)=0$.

\section{Solving the Equation of Motion}

\subsection{Solving the wave equation}
We study solutions of the one-dimensional wave equation which is the
homogeneous part of the inhomogeneous wave equation
\eqref{eq:eomPertFirstOrderHom}.  The spatial Dirichlet boundary
conditions tell us that the solution is a Fourier series with basis
functions $\sin m(x+\pi/2)$ with $m=1,2,3,\ldots$.  The solution to
the wave equation is
\begin{multline}
	\eta(\tau,x) = 
	\frac{2}{\pi} \sum_{n=1}^{\infty} \biggl[
	\biggl( A_{2n-1} \cos((2n-1)\tau) \\
	+ B_{2n-1} \sin((2n-1)\tau) \biggr)
	\cos((2n-1)x )\biggr]  \\
	 + \frac{2}{\pi} \sum_{n=1}^{\infty} \qty[ \biggl( A_{2n}
	\sin(2n\tau) - B_{2n} \cos(2n\tau) \biggr) \sin(2nx)],
	\label{eq:etaSeriesSoln}
\end{multline}
where the Fourier coefficients $A_{m}$ and $B_{m}$ are real numbers.
We see that the allowed frequencies $\omega$ are as expected all real,
and they are given by $\omega = 1,2,3,4,\dotsc$.

\subsection{Solving the inhomogeneous equation}

Equation \eqref{eq:eomPertFirstOrderHom} is an inhomogeneous linear
equation in $\varphi_{1}$.  We know that to find the solution, we need
the sum of a particular solution of the inhomogeneous equation and an
appropriate solution of the corresponding homogeneous equation.  We
therefore write the solution in the form $\varphi_{1}(\tau,x) = \xi(x)
+ \eta(\tau,x)$.
Because we have a $\tau$-independent source term, we take the
particular solution $\xi(x)$ to be a $\tau$-independent solution of
the inhomogeneous equation
\begin{align}
\dv[2]{\xi}{x} &= \frac{1}{2} \tilde{\mu}^{2} \sec[2](x) \varphi_{0}
\qty(\varphi_{0}^{2} - 1) \,, \label{eq:odeXi}
\end{align} 
while $\eta(\tau,x)$ is the general solution to the homogeneous
equation
\begin{align}
-\pdv[2]{\eta}{\tau} + \pdv[2]{\eta}{x} &= 0 \,.
\label{eq:pdeEta}
\end{align}
given by  \eqref{eq:etaSeriesSoln}.

\begin{figure}[tbp]
		\centering
		\includegraphics[width=\linewidth]{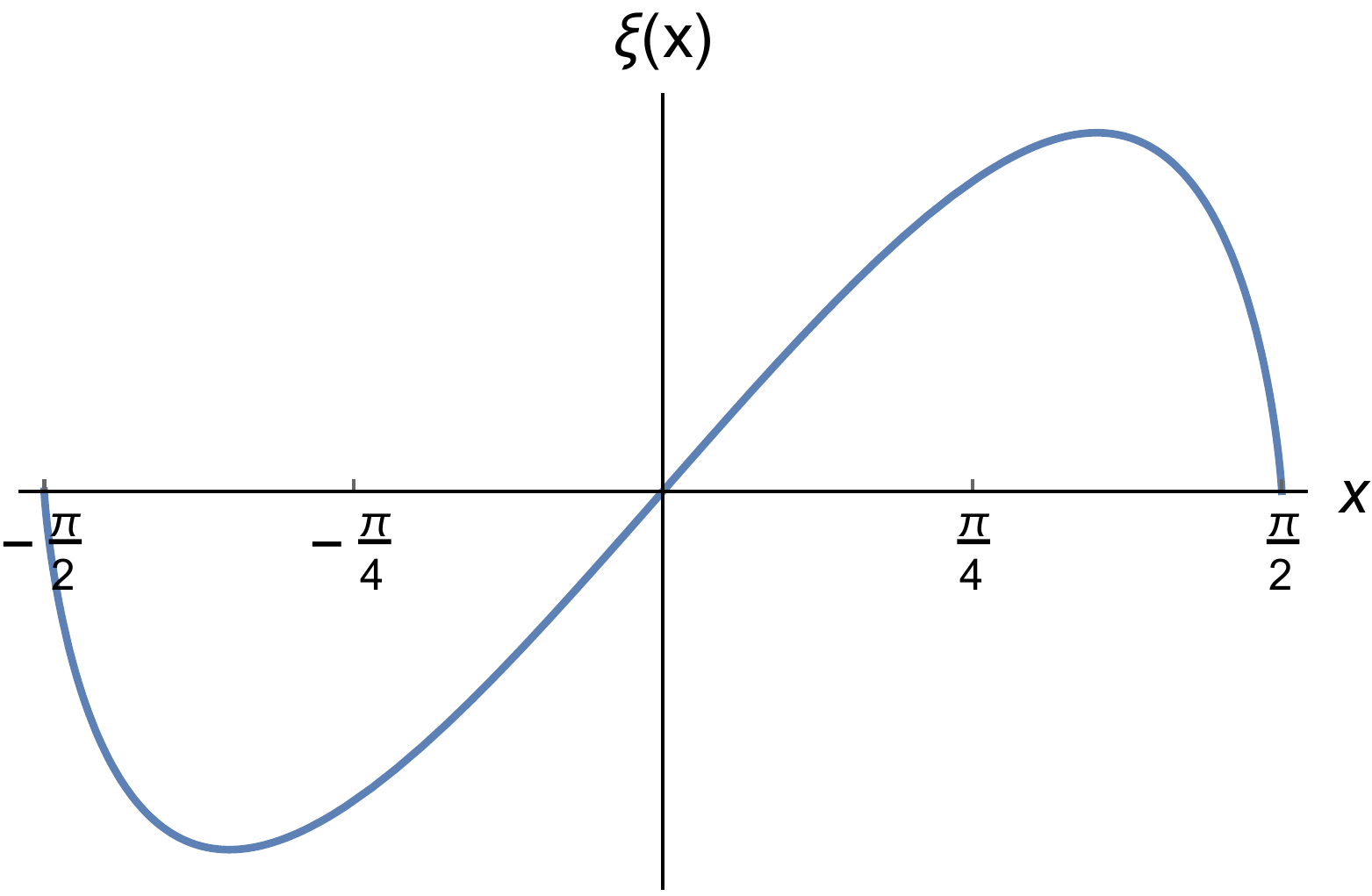}
	\caption{Plot of the first order correction $\xi(x)$.}
	\label{fig:xiPlot}
\end{figure}

\begin{figure}[tbp]
		\centering
		\includegraphics[width=\linewidth]{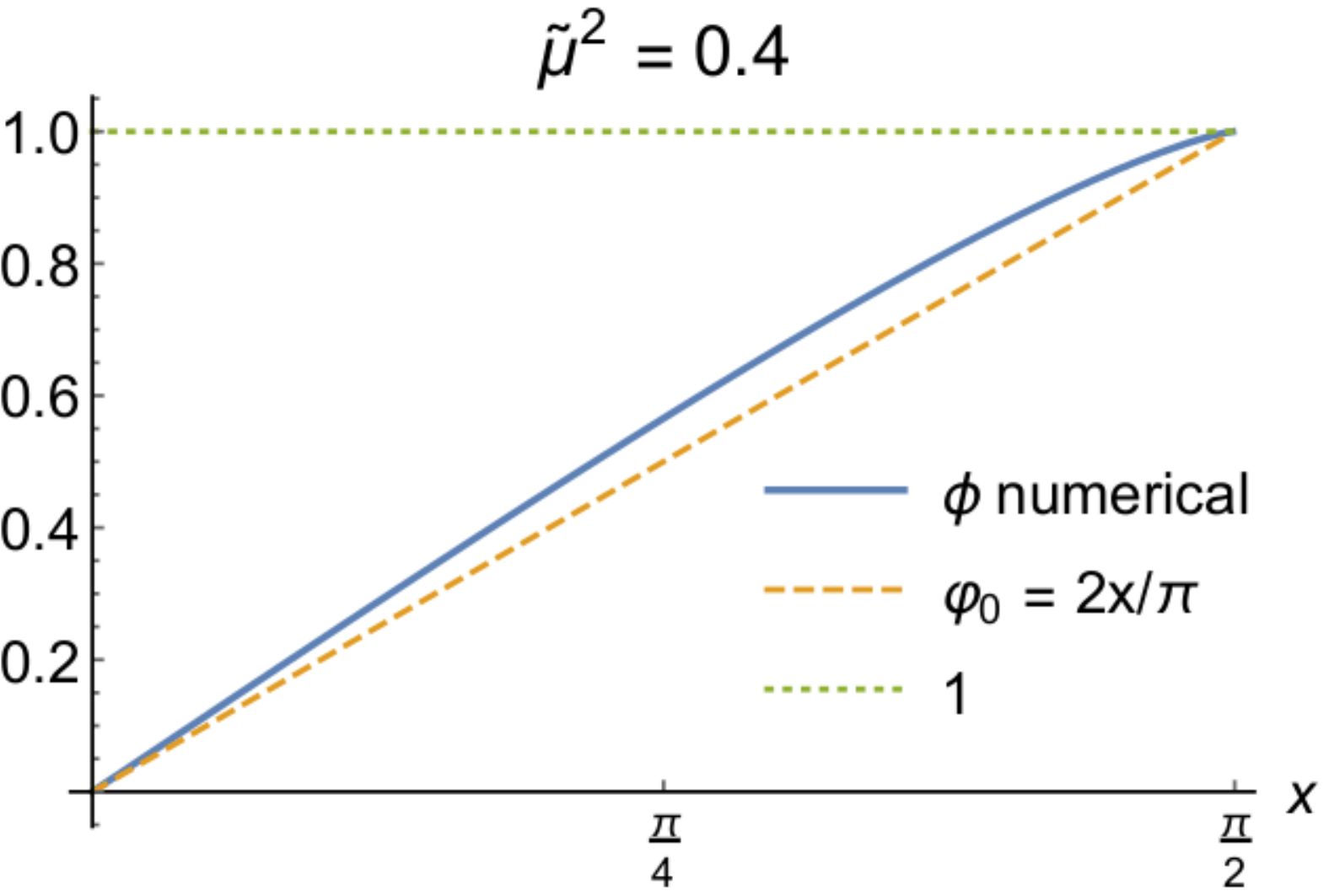}
	\caption{Plot of the kink given by the static numerical solution
	to the nonlinear ODE eq.~\eqref{eq:eomNoDimConformal} for a modest
	value of $\tilde{\mu}^{2}=0.4$, and a comparison to the zeroth
	order solution $\varphi_{0}$.  The function $\phi$ is odd and we
	only plot the interval $[0,\pi/2]$.}
	\label{fig:phiNumerical}
\end{figure}

Equation \eqref{eq:odeXi} is an ordinary differential equation in $x$.
The defect boundary conditions require that $\xi(x) \xrightarrow{x\to
\pm \pi/2} 0$.  If we substitute expression
\eqref{eq:dbpsSolnConformal} for $\varphi_{0}(x)$, the differential
equation then becomes:
\begin{align}
\dv[2]{\xi}{x} &= \frac{\tilde{\mu}^{2} (4x^{2} - 
	\pi^{2})}{\pi^{3}} x \sec[2](x)
	\label{eq:odeXiX}
\end{align}
The right hand side is an odd function and the solution should be an
odd function.  Integrating the equation twice yields an unwieldy
expression
\begin{align*}
\xi(x) &= \frac{\tilde{\mu }^2}{5 \pi ^3} \Big(-5 i \left(\pi 
^2-12 
x^2\right) \text{Li}_2\left(-e^{2 i x}\right)-90 x 
\text{Li}_3\left(-e^{2 i x}\right) \\
&\quad -60 i \text{Li}_4\left(-e^{2 i 
	x}\right) - 20 x^3 \log \left(1+e^{2 i x}\right)
	\nonumber \\
&\quad -5 i \pi 
^2x^2+90 x \,\zeta (3)+10 i \pi ^3 x+10 \pi ^2 x \log \left(1+e^{2 i 
	x}\right)
	\nonumber \\
	&\quad -5 \pi ^2 x \log 2
 -5\pi ^2 x \log (\cos x)-i \pi^4\Big) ,
\end{align*}
where $\text{Li}_{n}$ is the polylogarithm function of order $n$ and
$\zeta(3)\approx 1.20206$ is the  Riemann zeta function evaluated at
$3$. The solution $\varphi_{1}$ may be rewritten as
\begin{align}
\varphi_{1}(\tau,x) &= \xi(x) + \eta(\tau,x) 
	\nonumber \\
	&= \frac{\tilde{\mu }^2}{5 \pi ^3}
\biggl[5 \left(\pi ^2-12 x^2\right) \Im\left(\text{Li}_2\left(-e^{2 i
   x}\right)\right)     
   \nonumber \\
   &\quad -90 x \Re\left(\text{Li}_3\left(-e^{2 i
   x}\right)\right) +60 \Im\left(\text{Li}_4\left(-e^{2 i
   x}\right)\right)
   \nonumber \\
   &\quad 
   -5 x\left(4 x^2-\pi ^2\right) \log (2 \cos x)+90\zeta(3)\, x
    \biggr] 
	\nonumber \\
	&\qquad + \eta(\tau,x)
   \label{eq:xiSolnX}
\end{align}
We plot the function $\xi$ in figure~\ref{fig:xiPlot}.  In
figure~\ref{fig:phiNumerical}, we plot the static numerical solution
to \eqref{eq:eomNoDimConformal} and compare it to the zeroth order
solution.  The plots for $0 < \tilde{\mu}^{2} \lesssim 0.5$ all look
qualitatively similar.  You discover that $\phi'(0)$ is an increasing
function of $\tilde{\mu}^{2}$, and $\phi'(\pi/2)$ is a
decreasing\footnote{This is a consequence of the existence of a
Frobenius solution with positive exponent
$(\sqrt{1+4\tilde{\mu}^{2}\,}+1)/2$.} function of $\tilde{\mu}^{2}$.  Near $x=\pi/2$,
the general static solution of \eqref{eq:eomNoDimConformal} admits a
divergent Frobenius solution of the form
\begin{equation}
	\phi(x) \sim  \left(\frac{\pi}{2}
	-x\right)^{-(\sqrt{1+4\tilde{\mu}^{2}\,}-1)/2}\,.
	\label{eq:div-soln}
\end{equation}
In numerically solving the second order ODE, the initial condition
$\phi(0)=0$ is fixed, but care must be taken in choosing the value of
$\phi'(0)$ such that the divergent solution is avoided in order to
obtain the boundary condition $\phi(\pi/2)=1$.  We see from
\eqref{eq:div-soln} that the divergent behavior gets worse and
requires $\phi'(0)$ to be chosen with more precision as $\tilde{\mu}^{2}$
increases.  For this reason we stopped our numerical studies at
$\tilde{\mu}^{2} \approx 0.5$.

If we now analyze the second order perturbation equation of
motion~\eqref{eq:eomPertFirstOrder}, we note that in general the right
hand side of the equation is time dependent, therefore, the second
order correction $\varphi_{2}$ will be time dependent.

\section{General perturbation expansion of the equation of
motion\label{sec:GreensFn}}

Here we discuss general properties of our action and the equation of
motion.  Assume $\phi \in \mathbb{R}^{n}$, then our action is
schematically of the form
\begin{equation}
I(\phi) = \half\, q_{ij}\, \phi^{i}\phi^{j} - \epsilon U(\phi)\,
\end{equation}
where $q_{ij}$ is a real symmetric bilinear form. In our field
theoretic model, we have that $q= -\partial_{\tau}^{2} +
\partial_{x}^{2}$ is the d'Alembertian operator. The extrema of the
action gives the equation of motion
\begin{equation}
0 = q_{ij} \phi^{j} - \epsilon\, \frac{\partial U}{\partial
\phi^{i}}\,.
\end{equation}
We are interested in a power series expansion that leads to a
perturbative solution
\begin{equation}
\phi = \sum_{n=0}^{\infty}\epsilon^{n} \varphi_{n}\,.
\label{eq:phi-expansion}
\end{equation}
The order by order equation of motion is summarized by the power
series
\begin{align}
0 & = q\varphi _0  \nonumber \\
&\quad +\epsilon  \left(q\varphi _{1} -U'\left(\varphi
_0\right)\right)\nonumber \\
&\quad +\epsilon^{2} \left(q\varphi _2 -\varphi _1 U''\left(\varphi
_0\right)\right) \nonumber \\
   &\quad + \epsilon ^3 \left(q \varphi _3-\frac{1}{2} \varphi _1^2
   U^{(3)}\left(\varphi _0\right)-\varphi _2 U''\left(\varphi
   _0\right)\right) \nonumber \\
   &\quad + \epsilon ^4 \biggl(q \varphi _4-\frac{1}{6} \varphi _1^3
   U^{(4)}\left(\varphi _0\right)-\varphi _2 \varphi _1
   U^{(3)}\left(\varphi _0\right)
   \nonumber \\
   & \qquad -\varphi _3 U''\left(\varphi
   _0\right)\biggr)+ \order{\epsilon ^5}
  \label{eq:EOM-pert}
\end{align}
The static zeroth order solution $\varphi_{0}(\tau,x) = 2x/\pi$, see
\eqref{eq:dbpsSolnConformal}, already satisfies the spatial boundary
conditions $\phi(\tau,\pm \pi/2) = \pm 1$.  This means that all the
higher order terms satisfy vanishing Dirichlet spatial boundary
conditions $\varphi_{n}(\tau, \pm \pi/2)=0$ for $n \ge 1$.  An
important consequence is that if we write the $n$-th order
perturbative equation of motion as $q \varphi_{n} = s_{n}$ for $n \ge
2$, where $s_{n}$ is a source term that only contains lower order
field perturbations, then $s_{n}(\tau,x)=0$ at the spatial boundaries
$x = \pm \pi/2$.  With vanishing boundary conditions, the wave
equation, $q\psi=0$, has non-trivial solutions,
see~\eqref{eq:etaSeriesSoln}.  This means that the solution to
$q\varphi_{n}=s_{n}$ is not unique because of the non-trivial kernel
of $q$.  To solve the uniqueness issue we proceed as follows.  A
Greens' function $G$ satisfies $qG=I$, and the general solution to the
equation $q\psi=s$ may be written\footnote{ The $U$ terms all vanish
at $x = \pm \pi/2$, this means that the source is in the image of the
operator $q$.  This is necessary for the construction of a right
inverse.} as $\psi = \psi_{0} + Gs$ where $q\psi_{0}=0$.  The
uniqueness issue for solutions of $q\varphi_{n}=s_{n}$ may be resolved
by an appropriate choice of a cokernel for the operator $q$.  This is
equivalent to choosing a particular Greens' function\footnote{The
difference of two choices of Greens' function is solution to the wave
equation.}, \emph{i.e.}, a right partial inverse for $q$.  Well-known
choices for the d'Alembertian operator in Minkowski space are: the
advanced Greens' function, the retarded Greens' function, the Feynman
Greens' function, etc.  Once we have made a choice of cokernel, we can
uniquely specify the solution to $q\varphi_{n}=s_{n}$ as $\varphi_{n}=
Gs_{n}$ for $n \ge 2$.  The kernel of $q$ is included only in the first
term $\varphi_{1}$ which can be freely specified.  Once $\varphi_{1}$
is given then all the higher order corrections $\varphi_{n}$, for
$n\ge 2$, are uniquely determined by using \eqref{eq:EOM-pert}.  Said
differently, once a free massless wave oscillating in the background
of the zeroth order kink is specified then all higher order
corrections are completely determined.  Summarizing, we have
constructed a perturbative solution to the equation of motion.
\begin{remark} 
	In quantum mechanical time-independent perturbation theory, where
	you have a positive definite inner product in the Hilbert space,
	the perturbations are taken to be orthogonal to the zeroth order
	solution.  The argument we gave above is the generalization of
	this notion to our situation.
\end{remark}

In solving $q\psi=s$, we note that we can write two Fourier expansions
\begin{align*}
\psi(\tau,x) &= \int_{-\infty}^{\infty} \frac{\dd\omega}{2\pi}
\sum_{m=1}^{\infty} \tilde{\psi}_{m}(\omega)\; e^{-i\omega\tau} 
\sin\left[m\left(x +\frac{\pi}{2}\right)\right], \\
s(\tau,x) &= \int_{-\infty}^{\infty}
\frac{\dd\omega}{2\pi}\sum_{m=1}^{\infty} \tilde{s}_{m}(\omega)\;
e^{-i\omega\tau}  
\sin\left[m\left(x +\frac{\pi}{2}\right)\right] .
\end{align*}
Our choice of solution to the equation $q\psi=s$ is given by
\begin{align}
\psi(\tau,x) &= \int_{-\infty}^{\infty} \frac{\dd\omega}{2\pi}
\sum_{m=1}^{\infty} \frac{1}{\omega^{2} - m^{2}}\;
\tilde{s}_{m}(\omega)\; e^{-i\omega\tau} 
\nonumber \\
&\quad\times\; \sin\left[m\left(x +\frac{\pi}{2}\right)\right] 
\end{align}
The choice of Greens' function is determined by the poles and the
integration contour in the $\omega$ complex plane.  For our purposes
in this article, it is not necessary to specify the choice of Greens'
function explicitly.

\subsection{Classical Energy}

The conserved energy $E(\phi)$ is given by the expression
\begin{equation}
\frac{\rho\, E(\phi)}{\phi_{0}^{2}} = \int_{-\pi/2}^{+\pi/2} \left( \half
(\partial_{\tau}\phi)^{2} + \half (\partial_{x}\phi)^{2} +
\epsilon\; U(\phi)\right) \dd x\;,
\label{eq:def-E}
\end{equation}
where $U(\phi) = \frac{1}{8}\,\tilde{\mu}^{2}\,\sec[2](x)
\qty(\phi^{2} - 1)^{2}$.  The factor of the radius of curvature $\rho$
appears because the coordinates $(\tau,x)$ are dimensionless.  For an
odd static solution to \eqref{eq:eomNoDimConformal}, the expression
for the energy simplifies to
\begin{align}
	\frac{\rho\, E(\phi)_{\text{static}}}{\phi_{0}^{2}} &= 
	\phi'(\pi/2) 
	\nonumber \\
	&\quad + 2\int_{0}^{\pi/2} \left( U(\phi) - \half \phi\, 
	U'(\phi)\right) \dd{x}
	\label{eq:E-static-1}
\end{align}
where we used the exact static equation of motion $\phi''(x) =
U'(\phi)$, and here $U'(\phi) = \pdv*{U}{\phi}$.

\begin{figure}[tbp]
	\centering
	\includegraphics[width=\linewidth]{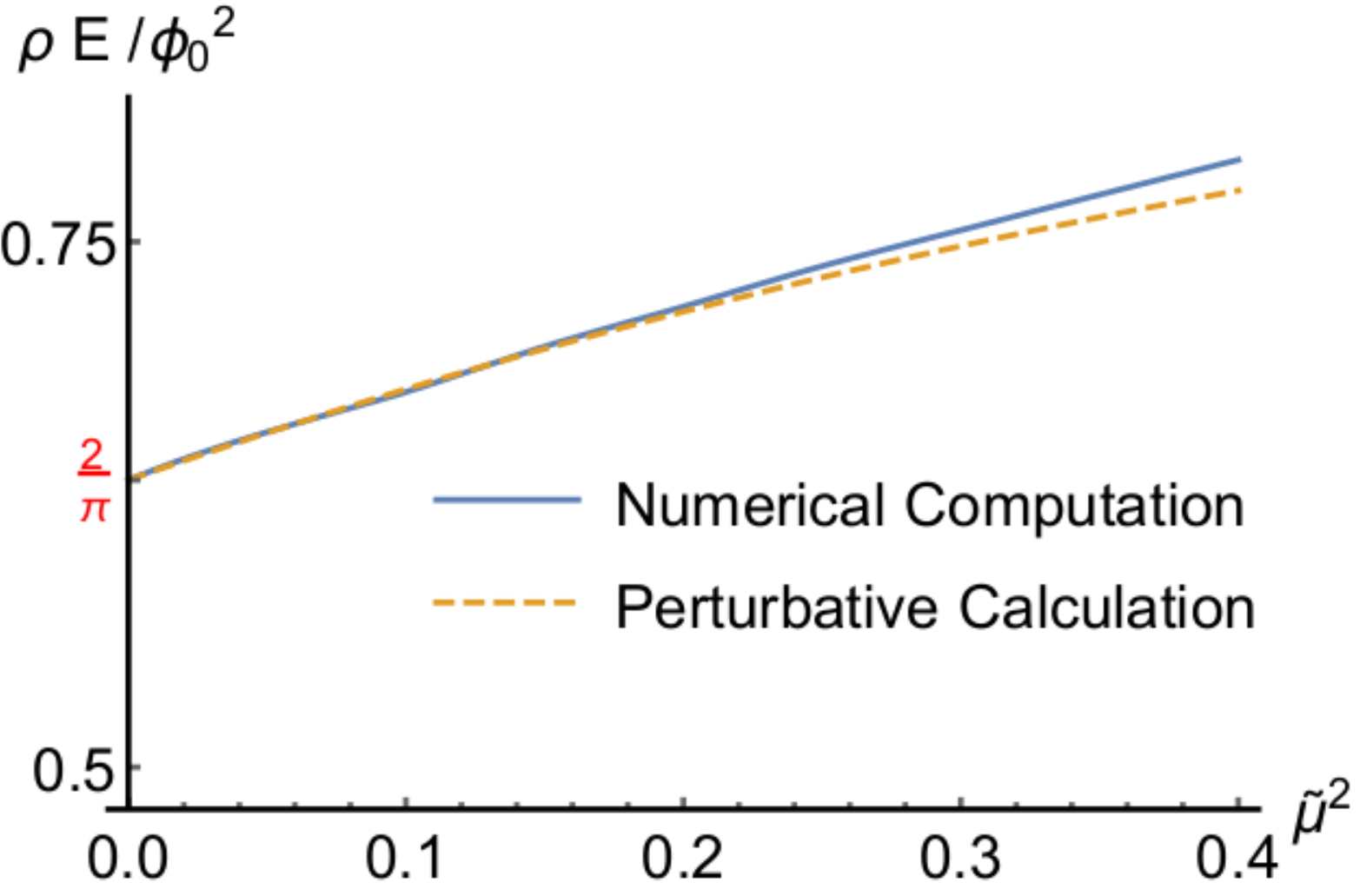}
	\caption{The energy of the static kink as a function of the
	parameter $\mu^{2}$ is computed by taking the numerical solution
	to the equations of motion and substituting into formula
	\eqref{eq:E-static-1}.  The numerical solution is compared to the
	perturbative expansion about the exact double BPS solution of the
	energy that may be extracted from \eqref{eq:energy}.  The
	dimensionless energy $\rho E/\phi_{0}^{2}$ of the double BPS kink
	is $2/\pi$.}
	\label{fig:energycomparison}
\end{figure}

For a non-static solution, we substitute perturbation expansion
\eqref{eq:phi-expansion} into the expression \eqref{eq:def-E} for
$E(\phi)$.  We use the fact that the spatial boundary conditions lead
to an identity $\int_{-\pi/2}^{\pi/2}(\partial_{x}\varphi_{n})\, \dd x
=0$ for $n\ge 1$, which leads to simplifications.  Substituting the
$\order{\epsilon^{2}}$ equation of motion found in
\eqref{eq:EOM-pert}, we conclude that
\begin{multline}
	\frac{\rho\, E(\phi)}{\phi_{0}^{2}} = \frac{2}{\pi} +\epsilon
	\int_{-\pi/2}^{+\pi/2} U\left(\varphi_{0}\right) \dd{x}
	\nonumber\\
	\quad + \epsilon^{2} \int_{-\pi/2}^{+\pi/2}
	\left(U'\left(\varphi_{0}\right) \varphi_{1}+\frac{1}{2}
	\qty(\partial_{\tau}\varphi_{1})^2 + \frac{1}{2}
	\qty(\partial_{x}\varphi_{1})^2\right) \dd{x} \nonumber \\
   \quad +\epsilon^{3} \int_{-\pi/2}^{+\pi/2} \biggl(\frac{1}{2}
   U''\left(\varphi_{0}\right)
   \varphi_{1}^{2}+U'\left(\varphi_{0}\right) \varphi_{2} \nonumber\\
   \qquad + \partial_{\tau}\varphi_{1} \partial_{\tau}\varphi_{2} 
   + \partial_{x}\varphi_{1} \partial_{x}\varphi_{2}
   \biggr) \dd{x}
   + \order{\epsilon^{4}}
\end{multline}
where $\varphi_{0}(\tau,x) = 2x/\pi$.    You can verify that the perturbation expansion of
$\phi$ to order~$\epsilon^{n}$ determines perturbation expansion of
$E(\phi)$ to order~$\epsilon^{n+1}$.  To see more explicitly the role
of the traveling wave $\eta(\tau,x)$, we substitute
$\varphi_{1}(\tau,x) = \xi(x) + \eta(\tau,x)$ into the expression 
above for $E(\phi)$ and we find

\begin{multline}
   \frac{\rho\, E(\phi)}{\phi_{0}^{2}} = \frac{2}{\pi} +\epsilon
   \int_{-\pi/2}^{+\pi/2} U\left(\varphi_{0}\right)  \dd{x} \nonumber \\
    +\epsilon^{2} \int_{-\pi/2}^{+\pi/2} \left( \half
   U'\left(\varphi_{0}\right) \xi (x) +
   \half\qty(\partial_{\tau}\eta)^{2} + \half \qty(\partial_{x}\eta)^{2} \right) \dd{x} \nonumber \\
    + \epsilon^{3} \int_{-\pi/2}^{+\pi/2} \biggl(\frac{1}{2}
   U''\left(\varphi_{0}\right) \xi(x)^2 + U''\left(\varphi_{0}\right)
   \xi (x) \eta (\tau ,x)
   \\
   +\frac{1}{2} U''\left(\varphi_{0}\right) \eta
   (\tau ,x)^2     + \partial_{\tau}\eta \partial_{\tau}\varphi_{2} 
   + \partial_{x}\eta \partial_{x}\varphi_{2} \biggr)  \dd{x} \nonumber \\
   +\epsilon^{4} \int_{-\pi/2}^{+\pi/2} \biggl( \sixth
  U^{(3)}\left(\varphi_{0}\right) \xi(x)^{3} +\half
  U^{(3)}\left(\varphi_{0}\right) \xi(x)^{2} \eta (\tau ,x) \nonumber\\
    + \half U^{(3)}\left(\varphi_{0}\right) \xi(x) \eta(\tau
   ,x)^{2}+ \sixth U^{(3)}\left(\varphi_{0}\right) \eta(\tau ,x)^{3}
   \nonumber \\
    + U''\left(\varphi_{0}\right) \xi(x) \varphi_{2}(\tau ,x) +
   U''\left(\varphi_{0}\right) \eta (\tau ,x) \varphi _2(\tau ,x)
   \nonumber\\
    + \half \qty(\partial_{x}\varphi_{2})^2
	+ \half\qty(\partial_{\tau}\varphi_{2})^2 \nonumber \\
    + \partial_{x}\eta \partial_{x}\varphi_{3} +
   \partial_{\tau}\eta \partial_{\tau}\varphi_{3} 
   \biggr)  \dd{x}  + \order{\epsilon^{5}}
\end{multline}
We can substitute into the above expression the explicit formulas for
$\varphi_{0}$ and $\xi$ obtained for our choice of $U$.  This
leads to some integrals that can either be computed explicitly or
numerically.  The result is
\begin{multline}
   \frac{\rho\, E(\phi)}{\phi_{0}^{2}} = \frac{2}{\pi} + \frac{12\,
   \zeta(3) }{\pi^{3}}\; \epsilon\, \tilde{\mu}^{2} 
   -0.339506\;\epsilon^{2} \,\tilde{\mu}^{4} \\
   + 0.589746\;\epsilon^{3} \,\tilde{\mu}^{6}
   +0.0829111\;\epsilon^{4} \,\tilde{\mu}^{8}
     \\
   +\epsilon^{2} \int_{-\pi/2}^{+\pi/2} \left(
   \half\qty(\partial_{\tau}\eta)^{2} + \half
   \qty(\partial_{x}\eta)^{2} \right) \dd{x}  \\
    + \epsilon^{3} \int_{-\pi/2}^{+\pi/2} \biggl(
   U''\left(\varphi_{0}\right) \xi (x) \eta (\tau ,x)+\frac{1}{2}
   U''\left(\varphi_{0}\right) \eta (\tau ,x)^2  \\
    + \partial_{\tau}\eta \partial_{\tau}\varphi_{2} +
   \partial_{x}\eta \partial_{x}\varphi_{2} \biggr) \dd{x} 
   \\
   +\epsilon^{4} \int_{-\pi/2}^{+\pi/2} \biggl( \half
  U^{(3)}\left(\varphi_{0}\right) \xi(x)^{2} \eta (\tau ,x) \\
    + \half U^{(3)}\left(\varphi_{0}\right) \xi(x) \eta(\tau
   ,x)^{2}+ \sixth U^{(3)}\left(\varphi_{0}\right) \eta(\tau ,x)^{3}
    \\
    + U''\left(\varphi_{0}\right) \xi(x) \varphi_{2}(\tau ,x) +
   U''\left(\varphi_{0}\right) \eta (\tau ,x) \varphi _2(\tau ,x)
   \\
    + \half \qty(\partial_{x}\varphi_{2})^2+
   \half\qty(\partial_{\tau}\varphi_{2})^2  \\
    +
   \partial_{\tau}\eta \partial_{\tau}\varphi_{3}
   + \partial_{x}\eta \partial_{x}\varphi_{3}  
   \biggr)  \dd{x}  + \order{\epsilon^{5}}
   \label{eq:energy}
\end{multline}
where $12\, \zeta(3)/\pi^{3} \approx 0.465218$.  We reiterate that
once $\eta$ is specified, everything in the energy formula
\eqref{eq:energy} is known in principle.  The first few explicit
numerical terms (as listed above) are plotted against a calculation of
the classical energy obtained by our perturbative solution in
figure~\ref{fig:energycomparison} as a function of $\mu^{2}$.  The
explicitly computed terms are the perturbative expansion for the
energy of the static soliton solution to
equation~\eqref{eq:eomNoDimConformal}.

The mass of the zeroth order kink is $2 \phi_{0}^{2}/\pi \rho$.  There
are two distinct order $\epsilon^{2}$ corrections to the energy.
Firstly, we have the explicitly computable static soliton
contribution.  Secondly, we have the energy of a massless wave
background $\eta(\tau,x)$.  We see that the interaction of the soliton
with the wave background contributes to the energy at order
$\varepsilon^{3}$.

\section{Conclusion}

Starting with the double BPS solution \eqref{eq:dbpsSolnConformal},
and adding a small perturbation in the mass term to the equation of
motion \eqref{eq:eomNoDimConformal}, we have obtained the first-order
perturbative correction to the solution of the
equations of motion beginning.
This perturbative solution is an oscillating kink-like (codimension 1)
topological defect in $\AdS_{2}$.  In addition, in
section~\ref{sec:GreensFn}, we laid out the method for building this
perturbative expansion at each order and we see that all corrections of
$\order{\epsilon^{n}}$ for $n \ge 2$ are determined by the choice of
linear perturbation wave solution $\eta(\tau,x)$.  From  
\eqref{eq:EOM-pert}, 
we can see that all of the $\tau$ dependence in
the solution originates from $\eta(\tau,x)$.  All of the normal
modes in the linearized term have frequencies that are non-zero real
and therefore these modes are stable as expected because of the 
topological stability of the soliton.

\appendix

\section{Killing Vectors and the Linearized Equation of Motion}
In this section we review the role played by the Killing vectors for
the metric on $\AdS_{2}$ on the solutions to the equation of motion
for the (1,1) defect.  Let $\Phi=(\Phi^1,\Phi^2,\ldots)$ be a real
scalar field in a potential $U$.  In general, the equations of motion
for this field will be of the form
\begin{align}
\Box\Phi^{I} - \pdv{U}{\Phi^{I}} &= 0 \label{eq:genericEOM}
\end{align}
where $\Box = g^{\mu\nu}\nabla_{\mu}\nabla_{\nu}$ is the appropriate 
second-order differential d'Alembertian operator for $\AdS_{2}$.

Now suppose that $\vb{K}$ is a Killing vector for the metric 
$g_{\mu\nu}$. By definition, Lie differentiation in the direction of 
$\vb{K}$ leaves the metric unchanged, therefore, it commutes with the 
geometric d'Alembertian:
$\mathsterling_{\vb{K}}\Box = \Box \mathsterling_{\vb{K}}$. If we
take 
the Lie derivative along $\vb{K}$ of \eqref{eq:genericEOM}, we
obtain:
\begin{align}
0 &= \mathsterling_{\vb{K}} \Box \Phi^{I} - \mathsterling_{\vb{K}} 
\pdv{U}{\Phi^{I}}  \nonumber \\
&= \Box \mathsterling_{\vb{K}} \Phi^{I} -
\pdv[2]{U}{\Phi^{I}}{\Phi^{J}} 
\mathsterling_{\vb{K}}\Phi^{J}  \nonumber \\
&= \Box \qty(K^{\mu} \pdv{\Phi^{I}}{x^\mu}) - 
\pdv[2]{U}{\Phi^{I}}{\Phi^{J}} \qty(K^{\mu} \pdv{\Phi^{J}}{x^{\mu}}) 
 \label{eq:eomKillingCmp}
\end{align}
Next let $\Psi^{I} = K^{\mu} \pdv{\Phi^{I}}{x^\mu}$, then we see
that
\begin{align}
\Box \Psi^{I} - \pdv[2]{U}{\Phi^{I}}{\Phi^{J}} \Psi^{J} &= 0.
\label{eq:eomKillingVec}
\end{align}
For each Killing vector $\vb{K}$, we potentially obtain a non-trivial
solution $\Psi = K^{\mu}\partial_{\mu}\Phi$ to the linearization of
\eqref{eq:genericEOM}.

In flat space, some of these linearized solutions would correspond to
zero-frequency modes in the oscillating part of the solution, because
the translational Killing vectors do not depend on the timelike
coordinate.  In our case of the (1,1) defect in $\AdS_{2}$, we will
not find non-trivial zero-frequency modes.  Two of the Killing vectors
are no longer time independent, however, the modes generated are
non-zero frequency solutions of the linearized equation of motion.

We now carry out this computation explicitly, using 
\eqref{eq:metricConformal} as our metric. Killing's equations are 
given by
\begin{align}
\mathsterling_{\vb{K}} g_{ab} &= K^{c} \partial_{c} g_{ab} + g_{cb} 
\partial_{a} K^{c} + g_{ac}	\partial_{b} K^{c}
=0\label{eq:killingsEqs}
\end{align}
This leads to the following system of differential equations:
\begin{subequations}
	\label{eq:killingsEqSys}
	\begin{align}
	\pdv{K^{\tau}}{\tau} &= -\tan(x) K^{x} 
	\label{eq:killingsEqSys1} 
	\\
	\pdv{K^{x}}{\tau} &= \pdv{K^{\tau}}{x} 
	\label{eq:killingsEqSys2} \\
	\pdv{K^{x}}{x} &= -\tan(x) K^{x} \label{eq:killingsEqSys3}
	\end{align}
\end{subequations}
The solutions to these equations are three Killing vector fields:
\begin{align*}
K^{\mu}_{1} &= \mqty(1 \\ 0), \\ 
K^{\mu}_{2} &= 
\mqty(-\sin(\tau)\sin(x) \\ \cos(\tau)\cos(x)), 
\nonumber \\
 K^{\mu}_{3} &= 
\mqty(\cos(\tau)\sin(x) \\ \sin(\tau)\cos(x))
\end{align*}
The first vector is just $\vb{K}_{1} = \partial_{\tau}$ and 
corresponds to time translations. The second corresponds to space 
translations in flat space, which we can see from looking at the limit
of 
$\rho \to \infty$ (i.e., $\tau/\rho \to 0$ and $x/\rho \to 0$). 
In this limit, $\vb{K}_{2} \to \partial_{x}$. Additionally, we see 
that 
$\vb{K}_{3} \to x \partial_{\tau} + \tau \partial_{x}$, which is 
the generator of boosts in the $\tau x$-plane in flat space.

If we differentiate a known solution to our equation of motion in
these directions, we should still have a solution.  For our known
solution, we take \eqref{eq:dbpsSolnConformal}.  Since this is
time-independent, $\vb{K}_{1}\varphi(x) = 0$ leads to a trivial
solution to the equation of motion \eqref{eq:eomNoDimConformal}.
Differentiating using the other Killing vectors produces
\begin{equation}
\begin{split}
\vb{K}_{2}\varphi(x) &= \frac{2}{\pi} \cos(\tau)\cos(x), \\
 \vb{K}_{3} \varphi(x) &= \frac{2}{\pi} \sin(\tau)\cos(x)
\end{split}
\label{eq:dbpsInKillingDir}
\end{equation}
Looking back at the oscillating part of our perturbative solution 
\eqref{eq:etaSeriesSoln}, we can see that the $n=1$ mode in the
first summand corresponds to the 
term
\begin{equation*}
\frac{2}{\pi} \qty\Big( A_{1} \cos(\tau) + B_{1} \sin(\tau)) \cos(x)
= A_{1}\,\vb{K}_{2}\varphi(x) + B_{1}\,\vb{K}_{3}\varphi(x)
\end{equation*}
As we can see, the functions in \eqref{eq:dbpsInKillingDir}
correspond 
to modes of frequency $1$ in the perturbative solution  given by
\eqref{eq:etaSeriesSoln}.



\begin{thebibliography}{1}
\providecommand{\url}[1]{{#1}}
\providecommand{\urlprefix}{URL }
\expandafter\ifx\csname urlstyle\endcsname\relax
  \providecommand{\doi}[1]{DOI \discretionary{}{}{}#1}\else
  \providecommand{\doi}{DOI \discretionary{}{}{}\begingroup
  \urlstyle{rm}\Url}\fi

\bibitem{alvarezHaddad2018}
O.~Alvarez, M.~Haddad, Journal of High Energy Physics \textbf{2018}(3), 12
  (2018).
\newblock \doi{10.1007/jhep03(2018)012}.
\newblock \urlprefix\url{https://doi.org/10.1007/JHEP03(2018)012}

\bibitem{bogomolny1976}
E.B. Bogomolny, Soviet Journal of Nuclear Physics \textbf{24}, 449 (1976)

\bibitem{prasadSommerfield1975}
M.K. Prasad, C.M. Sommerfield, Physical Review Letters \textbf{35}(12), 760
  (1975).
\newblock \doi{10.1103/PhysRevLett.35.760}.
\newblock \urlprefix\url{https://doi.org/10.1103/PhysRevLett.35.760}

\bibitem{lugo1999}
A.R. Lugo, F.A. Schaposnik, Physics Letters B \textbf{467}(1-2), 43 (1999).
\newblock \doi{10.1016/S0370-2693(99)01178-8}.
\newblock \urlprefix\url{https://doi.org/10.1016/S0370-2693(99)01178-8}

\bibitem{lugo2000}
A.R. Lugo, E.F. Moreno, F.A. Schaposnik, Physics Letters B \textbf{473}(1-2),
  35 (2000).
\newblock \doi{10.1016/S0370-2693(99)01481-1}.
\newblock \urlprefix\url{https://doi.org/10.1016/S0370-2693(99)01481-1}

\bibitem{ivanova2017}
T.A. Ivanova, O.~Lechtenfeld, A.D. Popov, Journal of High Energy Physics
  \textbf{2017}(11), 17 (2017).
\newblock \doi{10.1007/jhep11(2017)017}.
\newblock \urlprefix\url{https://doi.org/10.1007/JHEP11(2017)017}

\bibitem{coleman1985}
S.~Coleman, \emph{Aspects of Symmetry: Selected Erice Lectures} (Cambridge
  University Press, Cambridge, 1985).
\newblock \doi{10.1017/CBO9780511565045}
\newblock \urlprefix\url{https://doi.org/10.1017/CBO9780511565045}.


\bibitem{mantonSutcliffe2004}
N.~Manton, P.~Sutcliffe, \emph{Topological Solitons}.
\newblock Cambridge Monographs on Mathematical Physics (Cambridge University
  Press, Cambridge, 2004).
\newblock \doi{10.1017/cbo9780511617034}.
\newblock \urlprefix\url{https://doi.org/10.1017/cbo9780511617034}.

\bibitem{magneticMonopoles}
A.~Rajantie,  Philosophical Transactions of the Royal Society A:
Mathematical, Physical and Engineering Sciences, \textbf{370} (2012).
\newblock \doi{10.1098/rsta.2011.0394}.
\newblock \urlprefix\url{https://doi.org/10.1098/rsta.2011.0394}.

\bibitem{Vachaspati:2015}
T.~Vachaspati, L.~Pogosian, D.A. Steer, Scholarpedia \textbf{10}(2), 31682
  (2015).
\newblock \doi{10.4249/scholarpedia.31682}.
\newblock \urlprefix\url{https://doi.org/10.4249/scholarpedia.31682}.
\newblock Revision \#150671

\end{thebibliography}

\end{document}